\documentclass[twocolumn]{aastex631}

\usepackage{amsmath,amssymb,amstext}
\usepackage{gensymb}
\usepackage{multirow}
\usepackage{rotating}
\usepackage{graphicx}
\usepackage{colortbl}

\graphicspath{{./}{Figures/}}

\shorttitle{TOI-421\,b JWST Transmission Spectrum}
\shortauthors{Davenport et al.}

\begin{document}
\title{TOI-421\,b: A Hot Sub-Neptune with a Haze-Free, Low Mean Molecular Weight Atmosphere}



\correspondingauthor{Brian Davenport}
\email{bdav@umd.edu}

\author[0009-0000-7367-5541]{Brian Davenport}
\affiliation{Department of Astronomy, University of Maryland, College Park, MD 20742, USA}

\author[0000-0002-1337-9051]{Eliza M.-R. Kempton}
\affiliation{Department of Astronomy, University of Maryland, College Park, MD 20742, USA}

\author[0000-0001-8236-5553]{Matthew C.\ Nixon}
\affiliation{Department of Astronomy, University of Maryland, College Park, MD 20742, USA}

\author[0000-0003-2775-653X]{Jegug Ih}
\affiliation{Space Telescope Science Institute, 3700 San Martin Drive, Baltimore, 21218, MD, USA}

\author[0000-0001-5727-4094]{Drake Deming}
\affiliation{Department of Astronomy, University of Maryland, College Park, MD 20742, USA}

\author[0000-0002-3263-2251]{Guangwei Fu}
\affiliation{Department of Physics and Astronomy, Johns Hopkins University, Baltimore, MD, USA}

\author[0000-0002-2739-1465]{E. M. May}
\affiliation{Johns Hopkins APL, Laurel, MD 20723, USA}

\author[0000-0003-4733-6532]{Jacob L. Bean}
\affiliation{Department of Astronomy \& Astrophysics, University of Chicago, Chicago, IL 60637, USA}

\author[0000-0002-8518-9601]{Peter Gao}
\affiliation{Earth \& Planets Laboratory, Carnegie Institution for Science, 5241 Broad Branch Road NW, Washington, DC 20015, USA}

\author[0000-0003-0638-3455]{Leslie Rogers}
\affiliation{Department of Astronomy \& Astrophysics, University of Chicago, Chicago, IL 60637, USA}

\author[0000-0002-2110-6694]{Matej Malik}

\begin{abstract}

Common features of sub-Neptunes atmospheres observed to date include signatures of aerosols at moderate equilibrium temperatures ($\sim$500--800 K), and a prevalence of high mean molecular weight atmospheres, perhaps indicating novel classes of planets such as water worlds. Here we present a \mbox{0.83--5 $\mu$m} JWST transmission spectrum of the sub-Neptune TOI-421\,b.  This planet is unique among previously observed counterparts in its high equilibrium temperature ($T_{eq} \approx 920$ K) and its Sun-like host star.  We find marked differences between the atmosphere of TOI-421\,b and those of sub-Neptunes previously characterized with JWST, which all orbit M stars.  Specifically, water features in the NIRISS/SOSS bandpass indicate a low mean molecular weight atmosphere consistent with solar metallicity, and no appreciable aerosol coverage.  Hints of SO$_2$ and CO (but not CO$_2$ or CH$_4$) also exist in our NIRSpec/G395M observations, but not at sufficient signal-to-noise to draw firm conclusions.  Our results support a picture in which sub-Neptunes hotter than $\sim$850 K do not form hydrocarbon hazes due to a lack of methane to photolyze.  TOI-421\,b additionally fits the paradigm of the radius valley for planets orbiting FGK stars being sculpted by mass loss processes, which would leave behind primordial atmospheres overlying rock/iron interiors.  Further observations of TOI-421\,b and similar hot sub-Neptunes will confirm whether haze-free atmospheres and low mean molecular weights are universal characteristics of such objects.


\end{abstract}

\section{Introduction}
One of the most exciting prospects in exoplanet science today is discovering the origin and makeup of sub-Neptunes, which are high-occurrence planets that have no solar system analogue \citep[e.g.,][]{borucki11, Fulton2017}. Observational atmospheric studies are thought to be a productive avenue to addressing these topics \citep[e.g.,][]{bean21}.  However, sub-Neptune atmospheres have long been difficult to characterize.  This is due to their smaller signal size compared to hot Jupiters, and also the prevalence of muted or absent features in their transmission spectra \citep[e.g.,][]{Kreidberg2014, Guo2020, Gao2023, Wallack2024, Piaulet2024, Schlawin2024}. The launch of JWST has allowed us to largely overcome the former concern, yet studying sub-Neptune atmospheres has remained challenging because of the latter. 

HST observations have shown that muted spectral features are especially common for sub-Neptunes with equilibrium temperatures ($T_{eq}$) ranging from 500\,--\,800~K \citep{Brande2024}. At these temperatures, methane is expected to be the dominant carbon-bearing species in hydrogen-dominated planetary atmospheres \citep[e.g.,][]{madhu11, moses13}. Methane is especially susceptible to photolysis, and the resulting hydrocarbon radicals tend to polymerize. As a consequence, theoretical studies predict that hydrogen-rich atmospheres at equilibrium temperatures $\lesssim 850$~K should form photochemical hazes, potentially similar in composition to hydrocarbon aerosols such as industrial soot or the tholin haze found on Titan \citep[e.g.,][]{zahnle09, Kempton2012, Morley2013, kawashima18}. 

Yet colder planets observed to date, i.e, K2-18\,b ($T_{eq}\approx 280$~K) and TOI-270\,d ($T_{eq}\approx 390$~K), have revealed \textit{featured} transmission spectra, free of aerosol obscuration \citep{tsiaras19,benneke19,mikalevans23}. This could be indicative of less efficient haze production at cooler temperatures, although further theoretical and observational work is required to support this claim.  The characterization of these clear-atmosphere sub-Neptunes with JWST has provided detections of multiple chemical species and precise constraints on their abundances for the first time among planets of this class \citep{Madhusudhan2023, Holmberg2024, Benneke2024}.  Some of the results have been unexpected, calling into question previous theories for how sub-Neptunes form and evolve.

The prevailing theory for sub-Neptune formation and evolution begins with sufficiently massive rock-and-iron cores that are able to accrete hydrogen-dominated envelopes composed of nebular gas.  These envelopes are thick enough, and the planets' surface gravities high enough, that the atmospheres can survive against complete erosion from subsequent escape processes (either core-powered or photoevaporative) during their evolution \citep{Owen2017, Gupta2019}.  This picture is supported by the existence of a radius gap between super-Earth and sub-Neptune planets identified from the Kepler mission \citep{Fulton2017}.  Interestingly, the  high atmospheric mean molecular weights (MMWs) implied by observations of certain sub-Neptunes, including TOI-270\,d \citep{Holmberg2024, Benneke2024}, GJ 9827\,d \citep{Piaulet2024}, and GJ 1214\,b \citep{kempton23, Gao2023, Schlawin2024}, are in tension with this picture of hydrogen-dominated primordial gas atmospheres.

The existence of high-MMW (or equivalently, high-metallicity) atmospheres imply different or new classes of sub-Neptune planets, such as water worlds (i.e., planets composed of tens of percent H$_2$O in bulk) or miscible-envelope planets (i.e., those with mixed H$_2$\,--\,H$_2$O outer envelopes), which defy our conventional expectations \citep[e.g.,][]{Kuchner2003, Leger2004, Luque2022, Nixon2024, Benneke2024}. Of note, however, is that the sub-Neptunes with high-MMW atmospheres discovered to-date all orbit M-dwarf hosts, whereas the Kepler radius valley and related formation theories apply to planets orbiting FGK stars.  This raises the question of the role of stellar environments and whether these planets truly represent the population of sub-Neptunes across stellar types, or reveal a formation or evolution pathway unique to M-dwarf hosts \citep[e.g.,][]{Ogihara2009, Burn2021}.

On the other side of the temperature scale from the cold planets, sub-Neptunes hotter than $\sim$850\,K are also expected to be haze free because methane, and thus the hydrocarbon precursors to haze formation, should be less abundant as carbon monoxide becomes the dominant carbon-bearing molecule \citep[e.g.,][]{fortney13}. In order to test this hypothesis, and to explore whether atmospheres of sub-Neptunes orbiting FGK stars are fundamentally different from those with M-dwarf hosts, we present a combined JWST NIRISS and NIRSpec transmission spectrum for the hot sub-Neptune TOI-421\,b.  This planet has a radius of 2.65\,$R_{\oplus}$, orbits a late-G host star, and notably has a high equilibrium temperature of $\sim$920~K  \citep{carleo20,krenn24}. TOI-421\,b thus falls in a region of sub-Neptune parameter space that is expected to be haze free but also that lacks previous atmospheric observations.  Furthermore, TOI-421\,b resides in the core of the sub-Neptune population discovered by the Kepler mission in terms of its size, orbital period, and host star.

In Section \ref{sec:methods}, we discuss our methods for analyzing the observational data and for the modeling we perform to fit the resulting spectrum. In Section \ref{sec:results}, we present our analysis of the transmission spectrum and the inferred atmospheric properties. We detail the implications of our results for this planet and for the sub-Neptune population in general in Section \ref{sec:discussion}, and we summarize our conclusions in Section \ref{sec:conclusion}.

\section{Methods}  \label{sec:methods}
\subsection{Observations}

We observed two transits of the sub-Neptune TOI-421\,b with JWST, one each with the NIRSpec \citep{Jakobsen2022} and NIRISS \citep{Doyon2023} instruments (program GO 1935; E.\ Kempton PI). The NIRSpec observation was performed using the G395M filter, covering 2.8\,--\,5.2 $\mu$m, which illuminates only the NRS1 detector, removing the wavelength gap and detector offset present when observing with G395H \citep{Moran2023}. Four groups per integration were chosen to prevent exceeding 75\% saturation for any pixel, with 3058 integrations for a total exposure time of 3.85 hours. The NIRISS observation was conducted in the SOSS mode, with the SUBSTRIP96 subarray, which covers \mbox{0.8\,--\,2.8 $\mu$m} and provides only the first order spectral trace to account for the brightness of TOI-421. Three groups per integration were chosen to also prevent exceeding 75\% saturation, with 1581 integrations for a total exposure time of 3.9 hours. 

\begin{figure*}
    \centering
    \includegraphics[width=\textwidth,trim={0 0 20mm 0},clip]{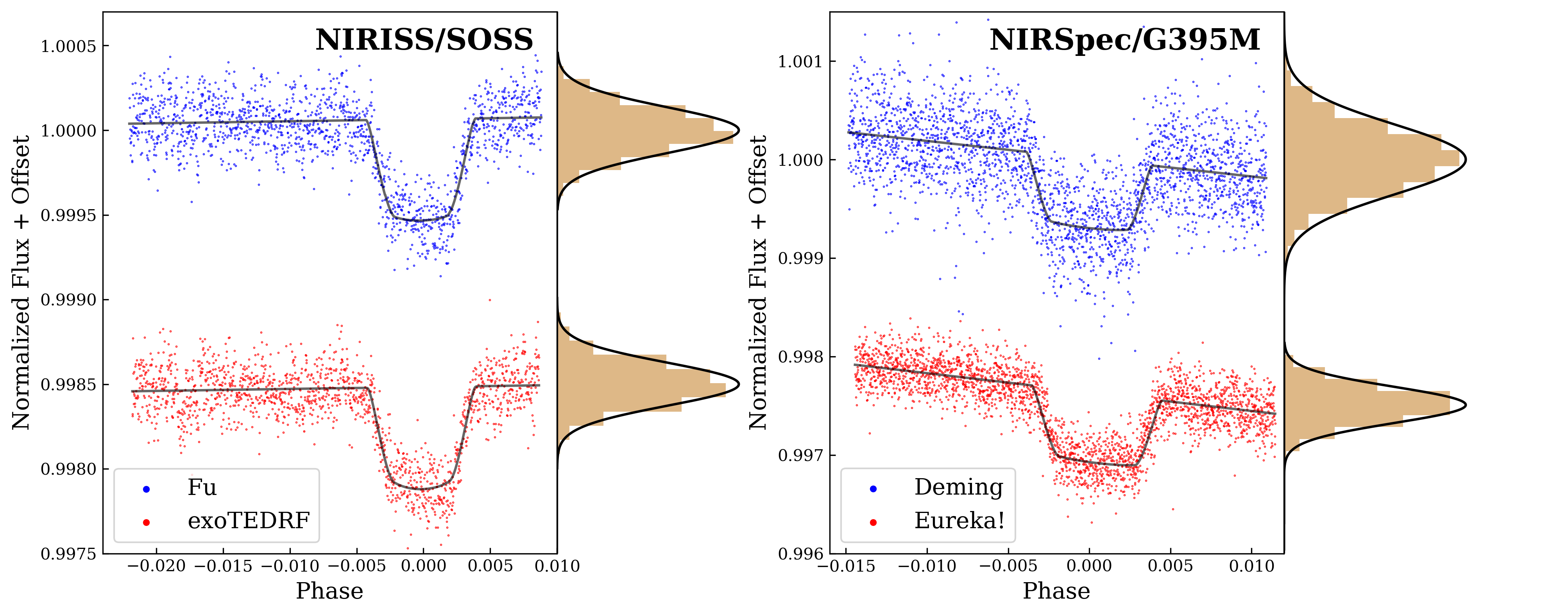}
    \caption{White light curves for NIRISS and NIRSpec observations along with model fits for each reduction method. On the right of each plot is a histogram showing residuals. The increased scatter in the Deming reduction relative to \texttt{Eureka!}\ for G395M is most likely due to different reduction choices, namely that the Deming reduction starts with \texttt{rateints} files and as a result does not run group-level background subtraction.}
    \label{fig:Light curves}
\end{figure*}

\subsection{Data Reduction}

We choose to conduct two independent reductions for each observation, as detailed below, in order to ensure that our results remain robust against the handling of systematics.  This choice is motivated by the fact that JWST is still a new observatory, and best practices for reducing JWST data are still evolving.   For example, for NIRISS/SOSS, it has been shown that misidentifying the true uncertainties in a given spectrum can impact atmospheric retrievals \citep{Holmberg2023}. Additionally, our NIRSpec observations utilize the G395M instrument mode, which to-date has not been widely used for exoplanet time series. For all reductions, planet and stellar parameters that we do not fit for are from \cite{krenn24}.

\subsubsection{NIRSpec: \texttt{Eureka!}}
We perform one full reduction of the NIRSpec observation with the \texttt{Eureka!}\ pipeline \citep{Bell2022} starting with the \texttt{uncal} data files. The first stage of \texttt{Eureka!}\  primarily serves as a wrapper around the \texttt{jwst} pipeline \citep{Bushouse2022} with the addition of a custom step for group-level background subtraction (GLBS) prior to ramp fitting which has been shown to improve the precision of NIRSpec light curves \citep[e.g.][]{Rustamkulov2023,Lustig-Yaeger2023}. Our GLBS routine first finds the trace and masks all pixels within 8 pixels on either side, then removes the median of all remaining pixels in a given column using an outlier rejection threshold of 5$\sigma$. This serves to correct for 1/$f$ noise which improves the accuracy of ramp fitting.

We run all the standard steps for stage 2 of the \texttt{jwst} pipeline through the \texttt{Eureka!}\ wrapper. \texttt{Eureka!}\ stage 3 performs spectral extraction. Wze optimal spectral extraction \citep{Horne1986}, where we optimize the aperture size and the background region, selecting 3 and 7 pixels respectively, by comparing the median absolute deviation of the resulting white light curves across a range of aperture and background values, while rejecting outliers over 6$\sigma$. The white light curves and fits for this and other reductions are shown in Figure \ref{fig:Light curves}. We use the resulting 1D time series spectra to create binned spectroscopic light curves at a resolution of $R=50$.

To create the transmission spectrum, we first fit the white light curve with Markov chain Monte Carlo (MCMC) sampling for planet radius, orbital period, center of transit, inclination, $a/R_*$, a linear relationship in time for out-of-transit flux, and a scatter multiplier to quantify white noise using \texttt{batman} \citep{Kreidberg2015}. Quadratic limb-darkening coefficients are also fixed at values calculated by \texttt{ExoTiC-LD} \citep{Grant2022} using the closest 1D Kurucz stellar model to TOI-421's parameters ($[\mathrm{M/H}] = 0.0$, $T_{\rm eff}$\,=\,5250~K, log$(g)=4.5$ in cgs units) \citep{Kurucz2004}. For the spectroscopic fits we fix the resulting orbital parameters of period (5.1967 days), center of transit (2460197.57258 HJD), inclination (85\degree), and $a/R_*$ (14.02) from the white light curve. We then fit for planet radius, an out-of-transit linear trend in time, and the scatter multiplier for each spectroscopic light curve, again using fixed \texttt{ExoTiC-LD} limb-darkening coefficients.

\subsubsection{NIRSpec: Deming}

As an independent comparison to the \texttt{Eureka!}\ reduction, D.\ Deming produced transit spectroscopy using the IDL programming language, and custom scripts written specifically for these data.  Using the \texttt{rateints} files, we extract spectra using a $\pm3$-row sum (not optimally weighted) centered on the peak row of the spectrum at each wavelength, finding the peak by fitting a Gaussian.  Background is defined and subtracted at each wavelength (column on the detector) as the median of 
all rows more than 3 pixels away from the peak row. We fit the white light curve using quadratic limb-darkening coefficients from \texttt{ExoCTK}, calculated using \texttt{ATLAS9} model atmospheres \citep{Kurucz2004}, and $T_{\rm eff}$\,/\,log($g$)\,/\,[M/H] = 5300\,/\,4.5\,/\,0.0.  The fit uses a linear ramp in time, and a gradient expansion algorithm to minimize $\chi^2$.  The white light fit includes refinement in orbital parameters that define the time of central transit and the impact parameter. When fitting individual wavelengths (via multi-variable linear regression), we force the limb-darkening coefficients to vary smoothly with wavelength, and we freeze the orbital parameters at their best-fit white light values. Some individual wavelengths have a stronger temporal ramp than does the white light curve, so we allow for quadratic ramps when deriving the transit spectrum.  After finding the transit depth at each wavelength (columns on the detector), we reject outliers using a $3\sigma$ clip, and we bin over wavelength on the same grid as the \texttt{Eureka!}\ reduction.  

\subsubsection{NIRISS: \texttt{exoTEDRF}}
We create one NIRISS transmission spectrum end-to-end with the \texttt{exoTEDRF} pipeline \citep{Radica2024, Radica2023,Feinstein2023}. The first several steps mirror the standard STScI JWST pipeline. For 1/$f$ noise subtraction, \texttt{exoTEDRF} first performs GLBS, in this case using the STScI SOSS model background for SUBSTRIP96. We then employ the pipeline's `scale-achromatic' method for group-level 1/$f$ noise correction, which finds a median value for each frame and does a column-by-column subtraction. The background is then re-added to the frames to allow for linearity fitting and flat fielding prior to final removal at the integration level. \texttt{exoTEDRF} uses the `edgetrigger' algorithm outlined in \cite{Radica2022} to locate the centroid of the spectral trace to allow for spectral extraction. For this next step we use a simple box extraction method, which assumes a constant-width trace in pixel space, versus more complex methods optimized for the overlapping three-trace frames of SUBSTRIP256. We optimize the spectral width by comparing the goodness-of-fit for the resulting white light curves from pixel widths of 27-36, finding that 32 pixels results in the best fit. During spectral extraction, we also test the inclusion of \texttt{PASTASOSS}, which updates the wavelength solution based on the actual position of the GR700XD grism's pupil wheel \citep{Baines2023a, Baines2023b}. As we find no discernible difference in resulting light curves or spectra, and since \texttt{PASTASOSS} has not previously been tested for SUBSTRIP96, we do not utilize the software for our final reduction.

We employ \texttt{exoTEDRF} for the light curve fitting, as well. For this, we fit the white light curve using \texttt{juliet} \citep{Espinoza2019}, incorporating \texttt{batman} \citep{Kreidberg2015} for the transit fitting, and \texttt{dynesty} \citep{Speagle2020} nested sampling for the fitting routine. At this stage, we fit for planet radius, orbital period, center of transit, inclination, $a/R_*$, a linear term in time for out-of-transit flux, and an additional error term to account for instrument jitter. Because of a noisy region early in the observation, 
and the allowance from a long baseline, we clip the first 620 integrations.  
Limb-darkening coefficients were also fixed at values calculated by \texttt{ExoTiC-LD} \citep{Grant2022} from 3D STAGGER-grid stellar models from the same stellar parameters used for the \texttt{Eureka!}\ reduction \citep{Magic2015}. Comparing the white light fits for orbital parameters to the \texttt{Eureka!}\ reduction, we find that they each agree to within $< 0.5$\%. From this result, we fix the values of period (5.1948 days), center of transit (2460254.740 HJD), inclination (86\degree), and $a/R_*$ (13.92) to perform the final fits for the spectroscopic light curves for planet radius and out-of-transit flux and create the transmission spectrum. 

\subsubsection{NIRISS: Fu}
G.\ Fu performed an independent reduction of the NIRISS/SOSS dataset. The overall reduction steps are similar to those used in \cite{Fu2022}. The STScI \texttt{jwst} pipeline is used to process the \texttt{uncal.fits} to generate the \texttt{darkcurrentstep.fits} file. GLBS is then performed before running the \texttt{RampFitStep} step to produce the final \texttt{rampfitstep.fits} dataset. Next, the spectra are extracted from each frame by cross-correlating the empirical spectral spread function along the wavelength direction and extracting with a 30-pixel width. Then each spectrum is summed in the vertical dispersion direction to form the white light and spectroscopic light curves. Next, we fit the white light curve with \texttt{batman} \citep{Kreidberg2015} with a systematics model consisting of a linear slope with time. The first of the two quadratic limb darkening coefficients is fixed and the second is free. The best-fit white light scaled semi-major axis ($a/R_*$) and inclination are then used for the spectroscopic light curves. The limb darkening parameters are fixed to the 3D Stagger-grid stellar atmosphere model interpolated to the best-fit stellar parameters \citep{Magic2015}. All spectroscopic light curves are fitted at the 1-pixel column level and the transit spectrum is then binned to the same resolution as the \texttt{exoTEDRF} reduction. 

\subsection{Atmospheric retrievals \label{sec:retrievals}}

\subsubsection{Aurora}

We constrain the atmospheric properties of TOI-421\,b using the \textit{Aurora} retrieval framework \citep{Welbanks2021,Nixon2024}, which combines a Bayesian parameter estimation scheme with an atmospheric forward model. The code is a generalization of the \textit{Aura} family of retrieval frameworks \citep{Pinhas2018,Welbanks2019a,Nixon2020,Nixon2022}.

\textit{Aurora} computes radiative transfer in the atmosphere of an exoplanet in hydrostatic equilibrium transiting its host star, assuming plane-parallel geometry. The forward model incorporates a variety of parametric temperature-pressure (TP) profiles, absorption from a range of chemical species, including collision-induced absorption (CIA), and a number of parametric treatments of cloud and haze opacity as well as stellar heterogeneity. These inputs are used to calculate a transmission spectrum at high resolution, which is convolved with the point spread function of the two instruments and binned to the resolution of the observations for comparison with real data.

        

For the retrievals shown in this work, we assume a H/He-dominated atmosphere in transmission geometry with solar relative abundances of H and He \citep{asplund2009}.  We include CIA due to H$_2$-H$_2$ and H$_2$-He \citep{richard2012}, as well as absorption due to the following chemical species with opacities taken from the referenced studies: H$_2$O \citep{Rothman2010}, CH$_4$ \citep{yurchenko2014}, NH$_3$ \citep{yurchenko2011}, HCN \citep{Barber2014}, CO \citep{Rothman2010}, CO$_2$ \citep{Rothman2010}, SO$_2$ \citep{Underwood2016} and H$_2$S \citep{Azzam2016}. Volume mixing ratios ($X_i$) for all species other than H and He are included as free parameters, with the remaining component of the atmosphere fixed as H/He in solar proportions \citep{asplund2009}. Priors for the volume mixing ratios are log-uniform, ranging from $10^{-12}$--$1$. If the sum of the mixing ratios of the chemical species exceeds 1, the likelihood is automatically set to zero to prevent unphysical solutions.

We assume an isothermal TP profile, with a uniform prior on atmospheric temperature $T_0$ from 200--1500~K. We found that our results were not sensitive to the choice of TP profile parameterization, and therefore opted for the simplest approach. We also retrieve the pressure at the white-light radius, $P_{\rm ref}$ (log-uniform prior from $10^{-8}$--$10^2$~bar). For clouds and hazes, we include a power-law haze plus a grey cloud deck, and allow for non-uniform cloud coverage \citep{Line2016}. We adopt standard priors for the cloud/haze parameters \citep[see e.g.,][]{Nixon2024}. We further allow for an absolute transit depth offset between the NIRISS/SOSS and NIRSpec/G395M spectra, adopting a Gaussian prior with a mean of 0 and a standard deviation of 100~ppm.  We additionally consider retrievals that include the effects of stellar heterogeneity (starspots and faculae) on the transmission spectrum, following the methods described in \citet{Rackham2017} and \citet{Pinhas2018}. We place a uniform prior of 0--0.5 on the heterogeneity covering fraction $\delta$, and a uniform prior of 0.5--1.5$\times T_{\rm eff}$ on the heterogeneity temperature, $T_{\rm het}$. 

We explore the model parameter space using the Nested Sampling algorithm \citep{Skilling2006}. Specifically, we use PyMultiNest \citep{Buchner2014}, a Python interface for MultiNest \citep{Feroz2008,Feroz2009}. Nested Sampling computes the Bayesian evidence for a given model, which allows us to compute the detection significance of a given molecule using Bayes factors between a model which includes that molecule and one which does not \citep{Trotta2008}.

\subsubsection{\textsc{platon}}

As a way of validating our retrieval results, we also run an independent set of retrievals using the code \textsc{platon} \citep{zhang19, zhang20,  ih21}.  The branch of \textsc{platon} used in this study is closest to version 5.1 and has been modified to allow for free retrievals and fractional aerosol coverage.  For a fair comparison with \textit{Aurora}, we use an identical list of included species and parameterizations for the temperature profile, offset between instruments, cloud and hazes, and stellar heterogeneity, as well as their priors.  The only difference between the two retrieval setups is that \textsc{platon} retrieves on the planet radius \citep[with a Gaussian prior derived from][]{krenn24} instead of retrieving on the reference pressure. As the primary effect of either of these parameterizations is only to establish a global offset for the transmission spectrum, we do not expect either choice to be consequential to the rest of the retrieval results \citep{welbanks19}.  We use opacities of model resolution $R$\,=\,$10^4$, which are then binned to the same resolution and wavelength grid as the data to calculate the goodness-of-fit.   We perform the same set of retrievals as \textit{Aurora}, both with and without stellar heterogeneity.  

In addition to the free retrievals, we perform two other tests to see how our conclusions hold up under different assumptions.  We perform an equilibrium chemistry retrieval on the full spectrum that replaces the individual molecular abundances with metallicity ($[\mathrm{M/H}]$) and carbon-to-oxygen ratio (C/O), in order to determine how the enforcement of chemical equilibrium impacts our results.  Also, upon finding that the most informative portion of TOI-421\,b's transmission spectrum is in the NIRISS bandpass (see Section~\ref{sec:results}), we perform a retrieval on only the NIRISS data using a simpler retrieval prescription.  This retrieval includes only an H$_2$O abundance and free MMW, along with the normal aerosol and offset parameters.  The inclusion of MMW as an independent parameter allows for constraining the mean molecular weight of the atmosphere with the least informative prior.

\begin{figure*}
    \centering
    \includegraphics[width=\textwidth]{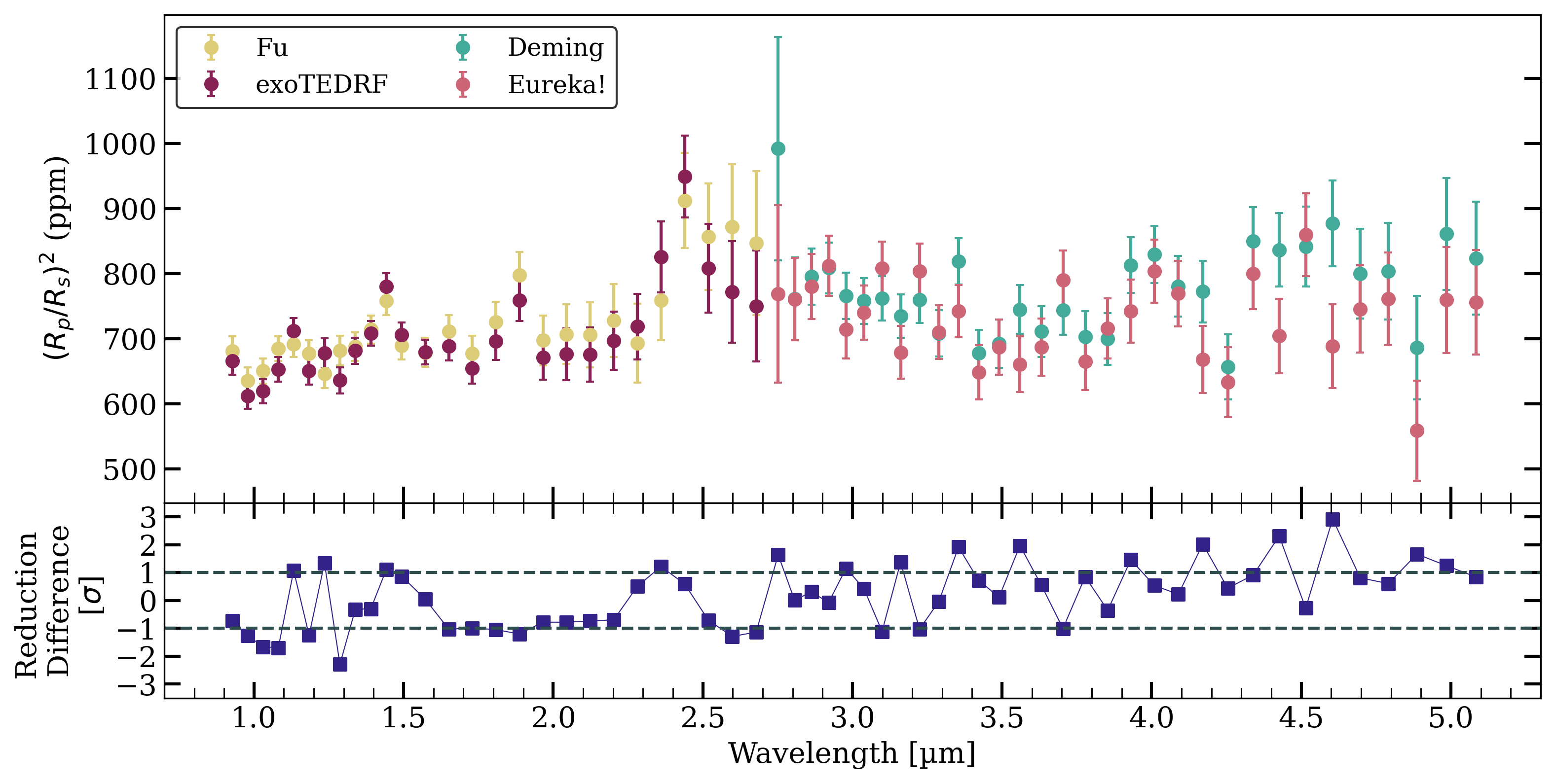}
    \caption{Comparison of transmission spectra between data reductions for the NIRISS/SOSS and NIRSpec/G395M observations, including the $\sigma$-difference between our modeled transit depths.}
    \label{fig:Transmission Spectrum}
\end{figure*}

\begin{figure*}
    \centering
    \includegraphics[width=\textwidth]{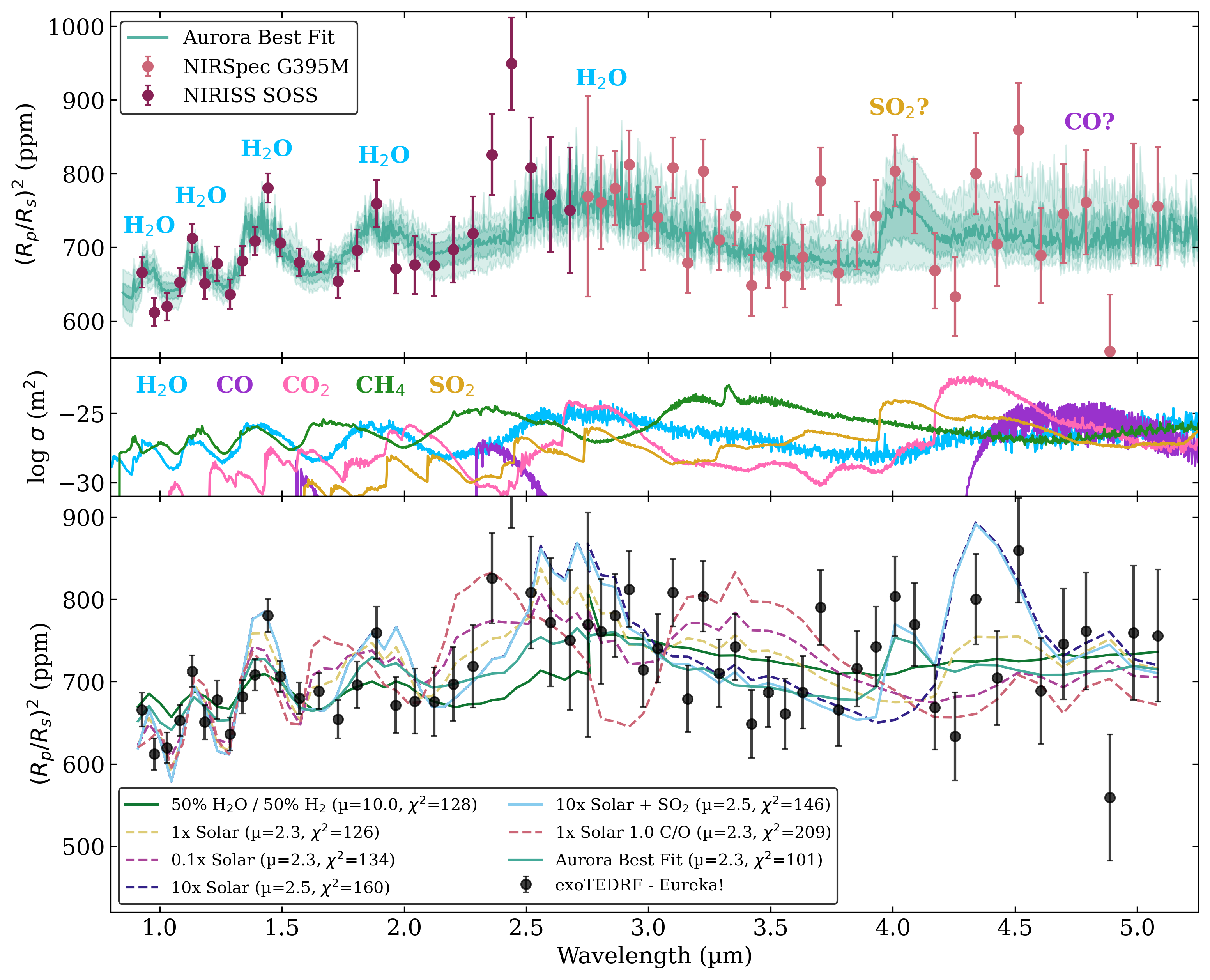}
    \caption{Comparison of the combined \texttt{exoTEDRF} + \texttt{Eureka!}\ transmission spectrum to \textit{Aurora} retrievals (top) and various \texttt{PLATON} forward models (bottom). The best-fit retrieval model is shown in the top panel with the darkness of the shading indicating the 1- and 2-$\sigma$ limits, and is re-plotted in the bottom panel for comparison against the other forward models. The middle panel shows the absorption cross-sections for key molecules, which demonstrates how the spectrum is a poor fit both for CH$_4$ and CO$_2$. The plotted forward models include different metallicity and C/O scenarios, as well as a water-rich composition, and the inclusion of disequilibrium SO$_2$. Forward models incorporating equilibrium chemistry are shown with dashed lines. The `10$\times$ Solar + SO$_2$' model has 10$^8$ times the equilibrium SO$_2$ abundance as its counterpart.  Offsets have been applied to the forward models at 2.8 $\mu$m to account for the best-fit offset between the NIRISS/SOSS and NIRSpec/G395M data.}
    \label{fig:Retrieval}
\end{figure*}

\section{Results}  \label{sec:results}
The resulting transmission spectrum from each of our reductions is presented in Figure \ref{fig:Transmission Spectrum}. 
We find that the individual analyses largely agree to within 1$\sigma$. A number of the spectral bins, however, do differ by greater amounts, especially in the 4\,--\,5 $\mu$m region, which influences the detection of molecules that absorb in this wavelength range. In order to determine the robustness of our results across our analyses, we run our atmospheric retrievals on both of the independently-reduced spectra (\texttt{ExoTEDRF} $+$ \texttt{Eureka!}\ and Fu $+$ Deming).

\begin{table*}
\hspace{-2.5cm}\begin{tabular}{cccccccccc}
\hline
\multicolumn{1}{|c|}{\multirow{2}{*}{\textbf{Retrieval}}} &
  \multicolumn{1}{c|}{\multirow{2}{*}{\textbf{Retrieval}}} &
  \multicolumn{1}{c|}{\multirow{2}{*}{\textbf{Data}}} &
  \multicolumn{7}{c|}{\textbf{Retrieved Value}} \\ \cline{4-10} 
\multicolumn{1}{|c|}{\multirow{2}{*}{\textbf{Assumption}}} &
  \multicolumn{1}{c|}{\multirow{2}{*}{\textbf{Software}}} &
  \multicolumn{1}{c|}{\multirow{2}{*}{\textbf{Pipeline}}} &
  \multicolumn{1}{c|}{\textbf{H$_2$O}} &
  \multicolumn{1}{c|}{\textbf{SO$_2$}} &
  \multicolumn{1}{c|}{\textbf{CO}} &
  \multicolumn{1}{c|}{\textbf{CH$_4$}} &
  \multicolumn{1}{c|}{\textbf{CO$_2$}} &
  \multicolumn{1}{c|}{\multirow{2}{*}{\textbf{MMW}}} &
  \multicolumn{1}{c|}{\textbf{Log}} \\ \cline{4-8}
\multicolumn{1}{|c|}{} &
  \multicolumn{1}{c|}{} &
  \multicolumn{1}{c|}{} &
  \multicolumn{5}{c|}{log abundance (detection significance {[}sigma{]})} &
  \multicolumn{1}{c|}{} &
  \multicolumn{1}{c|}{\textbf{P$_{\mathrm{cloud}}$}} \\ \hline
\multicolumn{1}{|c|}{\multirow{8}{*}{Baseline}} &
  \multicolumn{1}{c|}{\multirow{4}{*}{\textit{Aurora}}} &
  \multicolumn{1}{c|}{\multirow{2}{*}{e + E!}} &
  \multicolumn{1}{c|}{-3.81 $^{+1.32}_{-0.7}$} &
  \multicolumn{1}{c|}{-5.83 $^{+1.32}_{-2.16}$} &
  \multicolumn{1}{c|}{-6.42 $^{+2.99}_{-3.41}$} &
  \multicolumn{1}{c|}{\multirow{2}{*}{-8.39 $^{+1.91}_{-2.18}$}} &
  \multicolumn{1}{c|}{\multirow{2}{*}{-9.14 $^{+2.02}_{-1.71}$}} &
  \multicolumn{1}{c|}{\multirow{2}{*}{2.31 $^{+0.10}_{-0.01}$}} &
  \multicolumn{1}{c|}{\multirow{2}{*}{0.17 $^{+1.11}_{-2.07}$}}\\

\multicolumn{1}{|c|}{} &
  \multicolumn{1}{c|}{} &
  \multicolumn{1}{c|}{} &
  \multicolumn{1}{c|}{(4.51)} &
  \multicolumn{1}{c|}{(1.66)} &
  \multicolumn{1}{c|}{(1.28)} &
  \multicolumn{1}{c|}{} &
  \multicolumn{1}{c|}{} &
  \multicolumn{1}{c|}{} &
  \multicolumn{1}{c|}{} \\ \cline{3-10} 
\multicolumn{1}{|c|}{} &
  \multicolumn{1}{c|}{} &
  \multicolumn{1}{c|}{\multirow{2}{*}{F + D}} &
  \multicolumn{1}{c|}{-4.60 $^{+0.56}_{-0.47}$} &
  \multicolumn{1}{c|}{-5.85 $^{+0.63}_{-0.77}$} &
  \multicolumn{1}{c|}{-2.67 $^{+0.82}_{-0.92}$} &
  \multicolumn{1}{c|}{\multirow{2}{*}{-9.15 $^{+1.66}_{-1.89}$}} &
  \multicolumn{1}{c|}{\multirow{2}{*}{-9.29 $^{+1.85}_{-1.74}$}} &
  \multicolumn{1}{c|}{\multirow{2}{*}{2.36 $^{+0.28}_{-0.05}$}} &
  \multicolumn{1}{c|}{\multirow{2}{*}{0.21 $^{+1.15}_{-2.33}$}} \\
\multicolumn{1}{|c|}{} &
  \multicolumn{1}{c|}{} &
  \multicolumn{1}{c|}{} &
  \multicolumn{1}{c|}{(3.81)} &
  \multicolumn{1}{c|}{(2.35)} &
  \multicolumn{1}{c|}{(3.10)} &
  \multicolumn{1}{c|}{} &
  \multicolumn{1}{c|}{} &
  \multicolumn{1}{c|}{} &
  \multicolumn{1}{c|}{} \\ \cline{2-10} 
\multicolumn{1}{|c|}{} &
  \multicolumn{1}{c|}{\multirow{4}{*}{\textsc{platon}}} &
  \multicolumn{1}{c|}{\multirow{2}{*}{e + E!}} &
  \multicolumn{1}{c|}{-3.66 $^{+1.08}_{-0.78}$} &
  \multicolumn{1}{c|}{-5.37 $^{+1.20}_{-1.86}$} &
  \multicolumn{1}{c|}{-7.93 $^{+3.37}_{-2.80}$} &
  \multicolumn{1}{c|}{\multirow{2}{*}{-8.68 $^{+2.16}_{-2.22}$}} &
  \multicolumn{1}{c|}{\multirow{2}{*}{-9.20 $^{+1.92}_{-1.85}$}} &
  \multicolumn{1}{c|}{\multirow{2}{*}{2.31 $^{+0.08}_{-0.01}$}} &
  \multicolumn{1}{c|}{\multirow{2}{*}{0.83 $^{+1.44}_{-1.37}$}} \\
\multicolumn{1}{|c|}{} &
  \multicolumn{1}{c|}{} &
  \multicolumn{1}{c|}{} &
  \multicolumn{1}{c|}{(3.75)} &
  \multicolumn{1}{c|}{(not favored)} &
  \multicolumn{1}{c|}{(not favored)} &
  \multicolumn{1}{c|}{} &
  \multicolumn{1}{c|}{} &
  \multicolumn{1}{c|}{} &
  \multicolumn{1}{c|}{} \\ \cline{3-10} 
\multicolumn{1}{|c|}{} &
  \multicolumn{1}{c|}{} &
  \multicolumn{1}{c|}{\multirow{2}{*}{F + D}} &
  \multicolumn{1}{c|}{-4.43 $^{+0.85}_{-0.63}$} &
  \multicolumn{1}{c|}{-5.70 $^{+0.96}_{-1.14}$} &
  \multicolumn{1}{c|}{-3.30 $^{+1.42}_{-1.74}$} &
  \multicolumn{1}{c|}{\multirow{2}{*}{-8.72 $^{+1.96}_{-2.25}$}} &
  \multicolumn{1}{c|}{\multirow{2}{*}{-9.52 $^{+2.76}_{-1.72}$}} &
  \multicolumn{1}{c|}{\multirow{2}{*}{2.32 $^{+0.76}_{-0.02}$}} &
  \multicolumn{1}{c|}{\multirow{2}{*}{-0.11 $^{+1.33}_{-1.29}$}} \\
\multicolumn{1}{|c|}{} &
  \multicolumn{1}{c|}{} &
  \multicolumn{1}{c|}{} &
  \multicolumn{1}{c|}{(2.97)} &
  \multicolumn{1}{c|}{(not favored)} &
  \multicolumn{1}{c|}{(not favored)} &
  \multicolumn{1}{c|}{} &
  \multicolumn{1}{c|}{} &
  \multicolumn{1}{c|}{} &
  \multicolumn{1}{c|}{} \\ \hline
\multicolumn{1}{|c|}{} &
  \multicolumn{1}{c|}{\multirow{4}{*}{\textit{Aurora}}} &
  \multicolumn{1}{c|}{\multirow{2}{*}{e + E!}} &
  \multicolumn{1}{c|}{\multirow{2}{*}{-2.64 $^{+0.94}_{-0.92}$}} &
  \multicolumn{1}{c|}{\multirow{2}{*}{-4.93 $^{+1.24}_{-1.83}$}} &
  \multicolumn{1}{c|}{\multirow{2}{*}{-6.20 $^{+2.95}_{-3.48}$}} &
  \multicolumn{1}{c|}{\multirow{2}{*}{-8.52 $^{+2.12}_{-2.23}$}} &
  \multicolumn{1}{c|}{\multirow{2}{*}{-8.90 $^{+2.46}_{-1.98}$}} &
  \multicolumn{1}{c|}{\multirow{2}{*}{2.36 $^{+0.30}_{-0.05}$}} &
  \multicolumn{1}{c|}{\multirow{2}{*}{-0.29 $^{+1.46}_{-2.05}$}} \\
\multicolumn{1}{|c|}{} &
  \multicolumn{1}{c|}{} &
  \multicolumn{1}{c|}{} &
  \multicolumn{1}{c|}{} &
  \multicolumn{1}{c|}{} &
  \multicolumn{1}{c|}{} &
  \multicolumn{1}{c|}{} &
  \multicolumn{1}{c|}{} &
  \multicolumn{1}{c|}{} &
  \multicolumn{1}{c|}{} \\ \cline{3-10} 
\multicolumn{1}{|c|}{\multirow{3}{*}{Stellar}} &
  \multicolumn{1}{c|}{} &
  \multicolumn{1}{c|}{\multirow{2}{*}{F + D}} &
  \multicolumn{1}{c|}{\multirow{2}{*}{-4.30 $^{+0.74}_{-0.56}$}} &
  \multicolumn{1}{c|}{\multirow{2}{*}{-5.65 $^{+0.75}_{-0.78}$}} &
  \multicolumn{1}{c|}{\multirow{2}{*}{-2.42 $^{+0.74}_{-0.96}$}} &
  \multicolumn{1}{c|}{\multirow{2}{*}{-9.22 $^{+1.73}_{-1.77}$}} &
  \multicolumn{1}{c|}{\multirow{2}{*}{-8.93 $^{+2.15}_{-2.00}$}} &
  \multicolumn{1}{c|}{\multirow{2}{*}{2.42 $^{+0.47}_{-0.10}$}} &
  \multicolumn{1}{c|}{\multirow{2}{*}{-0.09 $^{+1.32}_{-1.99}$}} \\
\multicolumn{1}{|c|}{} &
  \multicolumn{1}{c|}{} &
  \multicolumn{1}{c|}{} &
  \multicolumn{1}{c|}{} &
  \multicolumn{1}{c|}{} &
  \multicolumn{1}{c|}{} &
  \multicolumn{1}{c|}{} &
  \multicolumn{1}{c|}{} &
  \multicolumn{1}{c|}{} &
  \multicolumn{1}{c|}{} \\ \cline{2-10} 
\multicolumn{1}{|c|}{\multirow{1}{*}{Heterogeneity}} &
  \multicolumn{1}{c|}{\multirow{4}{*}{\textsc{platon}}} &
  \multicolumn{1}{c|}{\multirow{2}{*}{e + E!}} &
  \multicolumn{1}{c|}{\multirow{2}{*}{-3.15 $^{+1.00}_{-0.90}$}} &
  \multicolumn{1}{c|}{\multirow{2}{*}{-4.90 $^{+1.20}_{-1.40}$}} &
  \multicolumn{1}{c|}{\multirow{2}{*}{-6.52 $^{+3.08}_{-3.67}$}} &
  \multicolumn{1}{c|}{\multirow{2}{*}{-8.38 $^{+1.78}_{-1.95}$}} &
  \multicolumn{1}{c|}{\multirow{2}{*}{-8.25 $^{+2.07}_{-2.17}$}} &
  \multicolumn{1}{c|}{\multirow{2}{*}{2.32 $^{+0.26}_{-0.02}$}} &
  \multicolumn{1}{c|}{\multirow{2}{*}{0.15 $^{+1.92}_{-3.68}$}} \\
\multicolumn{1}{|c|}{} &
  \multicolumn{1}{c|}{} &
  \multicolumn{1}{c|}{} &
  \multicolumn{1}{c|}{} &
  \multicolumn{1}{c|}{} &
  \multicolumn{1}{c|}{} &
  \multicolumn{1}{c|}{} &
  \multicolumn{1}{c|}{} &
  \multicolumn{1}{c|}{} &
  \multicolumn{1}{c|}{} \\ \cline{3-10} 
\multicolumn{1}{|c|}{} &
  \multicolumn{1}{c|}{} &
  \multicolumn{1}{c|}{\multirow{2}{*}{F + D}} &
  \multicolumn{1}{c|}{\multirow{2}{*}{-3.94 $^{+1.66}_{-0.76}$}} &
  \multicolumn{1}{c|}{\multirow{2}{*}{-5.93 $^{+1.67}_{-3.60}$}} &
  \multicolumn{1}{c|}{\multirow{2}{*}{-3.49 $^{+1.56}_{-4.46}$}} &
  \multicolumn{1}{c|}{\multirow{2}{*}{-8.91 $^{+1.59}_{-1.89}$}} &
  \multicolumn{1}{c|}{\multirow{2}{*}{-8.84 $^{+2.55}_{-2.00}$}} &
  \multicolumn{1}{c|}{\multirow{2}{*}{2.34 $^{+0.96}_{-0.04}$}} &
  \multicolumn{1}{c|}{\multirow{2}{*}{-0.25 $^{+1.77}_{-2.29}$}} \\
\multicolumn{1}{|c|}{} &
  \multicolumn{1}{c|}{} &
  \multicolumn{1}{c|}{} &
  \multicolumn{1}{c|}{} &
  \multicolumn{1}{c|}{} &
  \multicolumn{1}{c|}{} &
  \multicolumn{1}{c|}{} &
  \multicolumn{1}{c|}{} &
  \multicolumn{1}{c|}{} &
  \multicolumn{1}{c|}{} \\ \hline
\multicolumn{1}{|c|}{\multirow{4}{*}{NIRISS Only}} &
  \multicolumn{1}{c|}{\multirow{4}{*}{\textsc{platon}}} &
  \multicolumn{1}{c|}{\multirow{2}{*}{e + E!}} &
  \multicolumn{1}{c|}{\multirow{2}{*}{-4.46 $^{+0.63}_{-0.44}$}} &
  \multicolumn{1}{c|}{\multirow{2}{*}{---}} &
  \multicolumn{1}{c|}{\multirow{2}{*}{---}} &
  \multicolumn{1}{c|}{\multirow{2}{*}{---}} &
  \multicolumn{1}{c|}{\multirow{2}{*}{---}} &
  \multicolumn{1}{c|}{\multirow{2}{*}{2.37 $^{+1.03}_{-0.61}$}} &
  \multicolumn{1}{c|}{\multirow{2}{*}{1.24 $^{+1.16}_{-1.09}$}} \\
\multicolumn{1}{|c|}{} &
  \multicolumn{1}{c|}{} &
  \multicolumn{1}{c|}{} &
  \multicolumn{1}{c|}{} &
  \multicolumn{1}{c|}{} &
  \multicolumn{1}{c|}{} &
  \multicolumn{1}{c|}{} &
  \multicolumn{1}{c|}{} &
  \multicolumn{1}{c|}{} &
  \multicolumn{1}{c|}{} \\ \cline{3-10} 
\multicolumn{1}{|c|}{} &
  \multicolumn{1}{c|}{} &
  \multicolumn{1}{c|}{\multirow{2}{*}{F + D}} &
  \multicolumn{1}{c|}{\multirow{2}{*}{-4.95 $^{+0.68}_{-0.52}$}} &
  \multicolumn{1}{c|}{\multirow{2}{*}{---}} &
  \multicolumn{1}{c|}{\multirow{2}{*}{---}} &
  \multicolumn{1}{c|}{\multirow{2}{*}{---}} &
  \multicolumn{1}{c|}{\multirow{2}{*}{---}} &
  \multicolumn{1}{c|}{\multirow{2}{*}{2.58 $^{+1.40}_{-0.87}$}} &
  \multicolumn{1}{c|}{\multirow{2}{*}{1.27 $^{+1.20}_{-1.35}$}} \\
\multicolumn{1}{|c|}{} &
  \multicolumn{1}{c|}{} &
  \multicolumn{1}{c|}{} &
  \multicolumn{1}{c|}{} &
  \multicolumn{1}{c|}{} &
  \multicolumn{1}{c|}{} &
  \multicolumn{1}{c|}{} &
  \multicolumn{1}{c|}{} &
  \multicolumn{1}{c|}{} &
  \multicolumn{1}{c|}{} \\ \hline
 &
   &
   & 
   &
   &
   &
   &
   &
   &
   \\ \cline{7-8} 
 &
   & 
   & 
   &
   &\multicolumn{1}{c|}{}
   & \multicolumn{1}{c|}{\textbf{[M/H]}}
   & \multicolumn{1}{c|}{\textbf{C/O}}
   &
   &
\\ \hline
\multicolumn{1}{|c|}{\multirow{4}{*}{Equilibrium}} &
  \multicolumn{1}{c|}{\multirow{4}{*}{\textsc{platon}}} &
  \multicolumn{1}{c|}{\multirow{2}{*}{e + E!}} &
   &
   &
  \multicolumn{1}{c|}{} &
  \multicolumn{1}{c|}{\multirow{2}{*}{-0.35 $^{+0.65}_{-0.47}$}} &
  \multicolumn{1}{c|}{\multirow{2}{*}{0.46 $^{+0.21}_{-0.2 5}$}} &
  \multicolumn{1}{c|}{\multirow{2}{*}{2.31 $^{+0.03}_{-0.00 }$}} &
  \multicolumn{1}{c|}{\multirow{2}{*}{1.06 $^{+1.34}_{-1.43}$}} \\
\multicolumn{1}{|c|}{} &
  \multicolumn{1}{c|}{} &
  \multicolumn{1}{c|}{} &
   &
  &
  \multicolumn{1}{c|}{} &
  \multicolumn{1}{c|}{} &
  \multicolumn{1}{c|}{} &
  \multicolumn{1}{c|}{} &
  \multicolumn{1}{c|}{} \\ \cline{3-3}\cline{7-10} 
\multicolumn{1}{|c|}{} &
  \multicolumn{1}{c|}{} &
  \multicolumn{1}{c|}{\multirow{2}{*}{F + D}} &
   &
  &
  \multicolumn{1}{c|}{} &
  \multicolumn{1}{c|}{\multirow{2}{*}{1.28 $^{+1.15}_{-0.86}$}} &
  \multicolumn{1}{c|}{\multirow{2}{*}{0.75 $^{+0.08}_{-0.23}$}} &
  \multicolumn{1}{c|}{\multirow{2}{*}{2.66 $^{+3.60}_{-0.30}$}} &
  \multicolumn{1}{c|}{\multirow{2}{*}{-0.70 $^{+2.47}_{-0.72}$}} \\
\multicolumn{1}{|c|}{} &
  \multicolumn{1}{c|}{} &
  \multicolumn{1}{c|}{} &
  &
   &
  \multicolumn{1}{c|}{} &
  \multicolumn{1}{c|}{} &
  \multicolumn{1}{c|}{} &
  \multicolumn{1}{c|}{} &
  \multicolumn{1}{c|}{} \\ \hline
\end{tabular}
\caption{Retrieved and derived parameters from our full suite of retrievals.  The \texttt{exoTEDRF} + \texttt{Eureka!}\ and Fu + Deming retrievals are denoted by `e + E!' and `F + D', respectively.  MMW and cloud-top pressure ($P_{cloud}$) are given in units of amu and bar, respectively.  Detection significances are calculated for H$_2$O, SO$_2$, and CO for the baseline retrievals, which are our preferred retrievals, as described in the text. The detection significances themselves have a typical uncertainty of $\sim \pm 0.1$. In contrast to the \textit{Aurora} retrievals,  our \textsc{platon} retrievals prefer models that do not include either SO$_2$ or CO, based on their Bayesian evidence.  The \textsc{platon} equilibrium chemistry retrievals fit for metallicity (log, relative to solar) and carbon-to-oxygen ratio, rather than the abundances of individual species.}
\label{tab:results}
\end{table*}

Comparison of the data to our baseline \textit{Aurora} retrieval and to a set of \textsc{platon} forward models is shown in Figure \ref{fig:Retrieval}.  The retrieval plotted is considered our `baseline': it is a free retrieval allowing for an offset between instrument modes and no stellar heterogeneities.  To test the robustness of our retrieval results against various modeling assumptions, in Table \ref{tab:results} we also summarize the results of the full set of retrievals described in Section \ref{sec:retrievals}. 

From the combined set of retrievals and forward models, we find several consistent results.  First is the absence of clouds and hazes.  Both reductions disfavor a high-altitude gray cloud (i.e., $P_{cloud} \gtrsim 10\,$mbar in all of our baseline retrievals) and suggest a low aerosol coverage fraction at the terminator, with weak constraints on haze parameters suggesting that they do not contribute to the spectrum.  Second, in the absence of aerosols, we are able to constrain the atmospheric composition.  Specifically, we detect H$_2$O at near-solar abundances \mbox{($X_{\rm H_2O} \sim 10^{-3}$\,--\,$10^{-4}$)} across all of our baseline retrievals at $\gtrsim$ 3$\sigma$.  We find that the H$_2$O detection is driven primarily by the NIRISS data, with comparable constraints coming from our NIRISS-only \textsc{platon} retrieval as from an equivalent retrieval on the full spectrum.  We uniformly recover a slightly lower H$_2$O abundance from the Fu + Deming reduction compared to \texttt{exoTEDRF} + \texttt{Eureka!}, but the two reductions return consistent abundances to within 1.5$\sigma$.  Finally, we use the molecular abundances from our retrievals to derive the atmospheric MMW.  We find consistent values of 2.3\,--\,2.7 amu across all of our analyses, indicative of a H/He-dominated atmosphere. Most analyses favor MMW $<$ 2.4, which corresponds to at most very modest metal enrichment relative to solar (the MMW for solar composition is 2.3).

In addition to detecting H$_2$O, we obtain constraints on the abundances of other chemical species.  Both CH$_4$ and CO$_2$ are not detected.  We place stringent upper limits on their abundances at $<$1 ppm across all analyses, although we caution that the NIRSpec data are fairly noisy around 4.5 $\mu$m, which is where CO$_2$ would absorb most strongly.  We find marginal evidence for SO$_2$ absorption from both data reductions. The retrieved SO$_2$ abundances are consistent across our analyses with $X_{\rm SO_2} \sim 1$ ppm.  With our \textit{Aurora} retrievals, we note that the SO$_2$ abundance is more tightly constrained from the Fu+Deming reduction, and it is also detected at a slightly higher significance (2.4$\sigma$) compared to the \texttt{exoTEDRF} + \texttt{Eureka!} reduction ($1.7\sigma$).  In contrast, the \textsc{platon} retrievals do not favor the presence of SO$_2$ when the Bayesian evidence is compared against an equivalent set of retrievals run without this molecule, indicating the need for additional data to confirm its presence.

The main discrepancy between the two reductions can be found in the results for CO. Our retrievals consistently return a much higher (by several orders of magnitude) CO abundance with the Fu + Deming reduction than for \texttt{exoTEDRF} + \texttt{Eureka!}. This difference in interpretation can be attributed to the higher relative transit depths at 4.4\,--\,5 $\mu$m in the Deming reduction compared to the \texttt{Eureka!}\ reduction of the NIRSpec/G395M spectrum (see Figure \ref{fig:Transmission Spectrum}).  Our baseline \textit{Aurora} retrieval on the Fu + Deming reduction returns a 3.1$\sigma$ detection, whereas the equivalent \textsc{platon} retrieval does not find a statistical detection of CO, with a very weak preference of 1.8$\sigma$ for  a model with no CO, despite otherwise retrieving a high CO abundance.  As with SO$_2$, we conclude that the existing NIRSpec/G395M observations are not sufficient to definitively constrain the presence of CO. Additional observations over this wavelength range may be able to confirm these marginal detections and resolve existing discrepancies. 


\begin{figure*}
    \centering    \includegraphics[width=\textwidth]{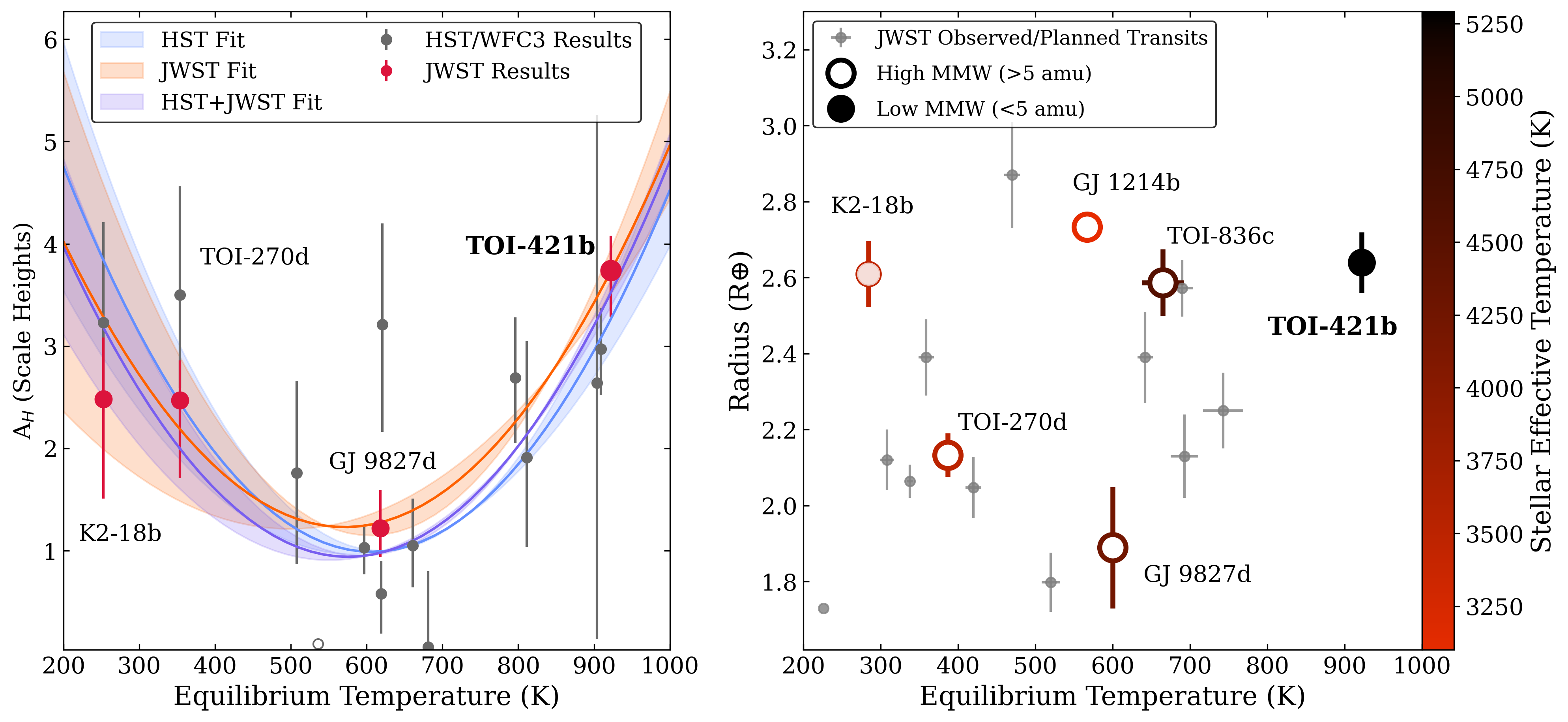}
    \caption{Placing TOI-421\,b in context of the characterized sub-Neptune population. Left: Amplitudes of the 1.4-$\mu$m transmission spectral feature in units of scale heights (A$_H$) for sub-Neptune exoplanets observed with HST-WFC3 (gray points) and JWST-NIRISS (red points).  GJ~1214\,b is included (open gray point) but not used for fitting, consistent with \cite{Brande2024}. Parabolic fits to the HST data only, JWST data only, and all planets versus $T_{eq}$ are shown in blue, orange, and purple, respectively. In the latter case, for the three planets observed with both HST and JWST, we fit only the JWST values in place of those from HST. All plotted values for HST data (gray points) are from \cite{Brande2024}.  JWST values (red points) are from our own retrievals on the JWST spectra, with the exception of GJ 9827\,d from P.\ Roy et al.\ (in prep.). In all cases, $A_H$ is calculated using the same methodology as in \citet{Brande2024}.
    Right: Planet size vs.\ equilibrium temperature for all JWST-observed sub-Neptunes (colored symbols) and planned JWST sub-Neptune observations (gray symbols).  Symbol color indicates the effective temperature of the stellar host, and filled (open) circles indicate derived MMW values less than (greater than) 5 amu.  For K2-18\,b, the true MMW is unknown due to the potential for water condensation in its cold atmosphere.  TOI-421\,b is unique among JWST-observed sub-Neptunes in having a high $T_{eq}$, orbiting a G-type host star, and having a low-MMW atmosphere.}
    \label{fig:Population}
\end{figure*}

Our results are largely consistent between the retrievals with and without stellar heterogeneity (see Table \ref{tab:results}). The only notable change is that the \texttt{exoTEDRF} + \texttt{Eureka!}\ reduction leads to a slightly higher H$_2$O abundance and a lower retrieved temperature when stellar heterogeneity is included.  Regardless, several factors lead us to conclude that the inclusion of stellar heterogeneity is not warranted in our retrievals.  First of all, our retrievals that incorporate stellar heterogeneity are disfavored compared to those that do not by \mbox{1.3\,--\,1.8$\sigma$}, for both retrieval codes and reductions.  Additionally, our stellar heterogeneity retrievals recover an implausibly large spot covering fraction ($\sim$10\%) and spot temperature ($\sim$6100~K), and an unphysically low atmospheric temperature (620\,--\,800~K) when compared against the planet's equilibrium temperature.   Finally, prior analyses of the G-type host star TOI-421 show that it is quiet and inactive based on its calcium H \& K activity index, rotation period, photometric variability, and age \citep{carleo20,krenn24}, making it unlikely that stellar activity is impacting our atmospheric inferences.

Our chemical equilibrium retrievals also produce results that are largely consistent with the free retrievals, with low MMWs and no aerosols being preferred.  However, the equilibrium retrieval on the Fu + Deming reduction returns a bimodal posterior, with a high-metallicity mode ([M/H] $\sim 10^2$) as one possibility. This retrieval is not statistically favored over the free retrievals by either of the goodness-of-fit or Bayesian evidence, and it returns an unphysically high atmospheric temperature of $\sim$1350~K.  Upon further inspection, we conclude that the interplay between trying to simultaneously fit for H$_2$O, CO, and atmospheric temperature, which are not decoupled under chemical equilibrium, is pushing this retrieval into an implausible region of parameter space.  We therefore disfavor this mode of the retrieval relative to our baseline free retrievals.

\section{Discussion}  \label{sec:discussion}

\subsection{Constraints on Atmospheric Composition and Aerosols}

\subsubsection{A Haze-Free Atmosphere}

Our retrievals suggest no presence of clouds or hazes at the pressures probed in these observations.  This is in line with both theoretical and empirical predictions that aerosol formation is unlikely at the $\sim$920~K equilibrium temperature of TOI-421\,b \citep[e.g.,][]{Morley2013, Morley2015, gao20, Brande2024, ashtari24}.  A key result of these previous studies is that hydrocarbon haze formation should be inefficient for T$_{eq} > 850$~K, and silicate clouds don't form high enough in the atmosphere to impact transmission spectra until yet hotter temperatures of $\sim$1300~K, leading to clear-sky conditions for planets with comparable irradiation to TOI-421\,b.  Our results provide observational evidence for this theory.

To place our no-aerosol findings in the context of existing atmospheric observations of Neptune- and sub-Neptune-size exoplanets, we add TOI-421\,b to the empirical trend in H$_2$O feature strength versus temperature found by \cite{Brande2024} (Figure~\ref{fig:Population}, left panel).  Fitting the strength of the 1.4-$\mu$m absorption feature versus equilibrium temperature yields a parabolic trend.  This is true regardless of whether the trendline is only fit to the HST data, JWST data, or both, as shown in Figure~\ref{fig:Population}.  Remarkably, TOI-421\,b falls exactly on the trendline established by HST data, perfectly in-line with predictions.  While the trend has previously been attributed to the onset of hydrocarbon haze production, predicted to peak around 600\,--\,800~K \citep{Morley2015}, the existence of compositional diversity among sub-Neptunes may further complicate interpretation.  For example, the smaller feature size for GJ 9827\,d is interpreted as being due to high MMW, rather than aerosols \citep{Piaulet2024}. 

\subsubsection{Atmospheric Composition \label{sec:composition}}
Based on our measured atmospheric abundances for TOI-421\,b, we calculate a median MMW between our baseline \textit{Aurora} and \textsc{platon} retrievals of 2.32 amu. This low value, along with our retrieved H$_2$O abundances and lack of a CO$_2$ detection, suggests TOI-421\,b hosts a hydrogen-dominated envelope at near-solar metallicity.  
The low recovered MMW of TOI-421\,b makes it unique among sub-Neptunes studied to-date with JWST. TOI-270\,d \citep[5.37;][]{Benneke2024}, GJ 9827\,d \citep[9.84;][]{Piaulet2024}, TOI-836\,c \citep[$>$6;][]{Wallack2024}, and GJ 1214\,b \citep[$>$15;][]{Gao2023}, all have constrained or lower-limit MMW values indicative of M/H\,$>$\,100$\times$ solar, and in some cases much higher. \mbox{K2-18\,b} \citep[3.1; calculated from][]{Madhusudhan2023}, the coldest observed sub-Neptune (T$_{eq} < 300$~K), is the only other such planet with MMW $<$\,5, but this planet's low temperature suggests the likelihood that a significant amount of water has condensed out of its atmosphere, suppressing the measured abundance of metals. 

TOI-421\,b thus adds to the compositional diversity of sub-Neptune planets. Its unique makeup may be attributable to its unique temperature and/or its stellar environment among sub-Neptunes studied with JWST to date. In addition to being the hottest planet of its type with a published JWST transmission spectrum by over 250~K, it is so far the only one orbiting a Sun-like star, rather than an M-dwarf. The right-hand panel of Figure \ref{fig:Population} demonstrates the unique stature of TOI-421\,b in composition, equilibrium temperature, and stellar host type. 

The lack of confident detections of carbon- and sulfur-bearing species in TOI-421\,b's atmosphere makes it difficult to constrain elemental abundance ratios such as C/O or S/H, which are otherwise informative tracers of formation or ongoing physical processes \citep[e.g.,][]{oberg11,madhu12,Seo2024, tsai23}.  For carbon, the lack of CH$_4$ in TOI-421\,b's transmission spectrum is consistent with its high equilibrium temperature and also with the absence of hazes, which could otherwise be formed as a by-product of methane photolysis in cooler atmospheres.  Similarly, the lack of CO$_2$ is consistent with TOI-421\,b's low MMW, as the equilibrium abundance of this molecule in hydrogen-dominated atmospheres is strongly correlated with metallicity \citep[e.g.,][]{lodders02,moses13}.  This leaves CO as the primary carbon-carrier for TOI-421\,b.  The poorly constrained abundance of this molecule arises from insufficient signal-to-noise in the NIRSpec data and the nature of the 4.5\,--5.0~$\mu$m CO bandhead being relatively broad and weak.  The resulting 1$\sigma$ upper limit on C/O from our \textit{Aurora} free retrievals  is 0.64 for the \texttt{ExoTEDRF} $+$ \texttt{Eureka!}\ reduction and 0.99 for the Deming + Fu reduction.  The higher value in the latter case results from higher inferred CO abundance.  We can therefore confidently conclude that C/O $< 1$ in TOI-421\,b's atmosphere, but additional measurements will be required to achieve tighter constraints.  For sulfur, the $\sim$1 ppm SO$_2$ abundance preferred by our baseline retrievals is consistent with photochemical models for planets with near-solar metallicity and comparable properties to TOI-421\,b \citep{mukherjee24}.

\subsection{Interior Modeling and Bulk Composition \label{sec:interior}}

The constraints placed on the atmospheric composition of TOI-421\,b from its transmission spectrum also allow us to better understand the planet's bulk composition and interior structure. We use the internal structure model presented in \citet{Nixon2021}, with updates described in \citet{Nixon2024_GJ1214b}, to model the interior of the planet. This model calculates a planet's radius for a given mass, composition and temperature structure. This is achieved by solving the equations of planetary structure (mass continuity, hydrostatic equilibrium) under the assumption of spherical symmetry for planets consisting of Fe, MgSiO$_3$, H$_2$O, and H/He. At temperatures close to the equilibrium temperature of TOI-421\,b, H$_2$O will be in vapor/supercritical phase, and will be well-mixed with H/He \citep{Gupta2024}. We therefore model the planet with a differentiated iron core and silicate mantle, and a mixed H/He/H$_2$O envelope, with the H$_2$O mass fraction chosen to give a mean molecular weight of 2.31~amu, consistent with the retrieved mean molecular weight. For the iron and silicate layers, we assume an Earth-like compositon of 1/3 iron, 2/3 silicates by mass, and we use a solar He mass fraction of 0.275 in the atmosphere.

We use the radiative-convective atmosphere model HELIOS \citep{Malik2017} to compute the thermal structure of the upper atmosphere of TOI-421\,b, which is then used as the upper boundary of the internal structure model. HELIOS solves the equation of radiative transfer to calculate a TP profile for a planet with a given composition. We use an atmospheric composition of 1$\times$ solar metallicity and C/O ratio, assuming chemical equilibrium, consistent with our constraints on the atmosphere from the transmission spectrum. The planet's intrinsic temperature $T_{\rm int}$ is required as an input to HELIOS, but is not constrained by our observations. We therefore consider two values of $T_{\rm int}$, 25~K and 50~K, covering a range of reasonable values expected for mature sub-Neptunes \citep{Lopez2014}. We assume an adiabatic temperature profile throughout the gaseous envelope and silicate mantle from the maximum pressure of the HELIOS models (1000 bar) down to the core-mantle boundary.  For the iron core, we assume an isothermal temperature structure, though the exact thermal structure of the core has been shown to have a negligible effect on resulting planetary radii \citep{Howe2014}.

\begin{figure}
\centering
\includegraphics[width=\columnwidth,trim={0 0 0 0},clip]{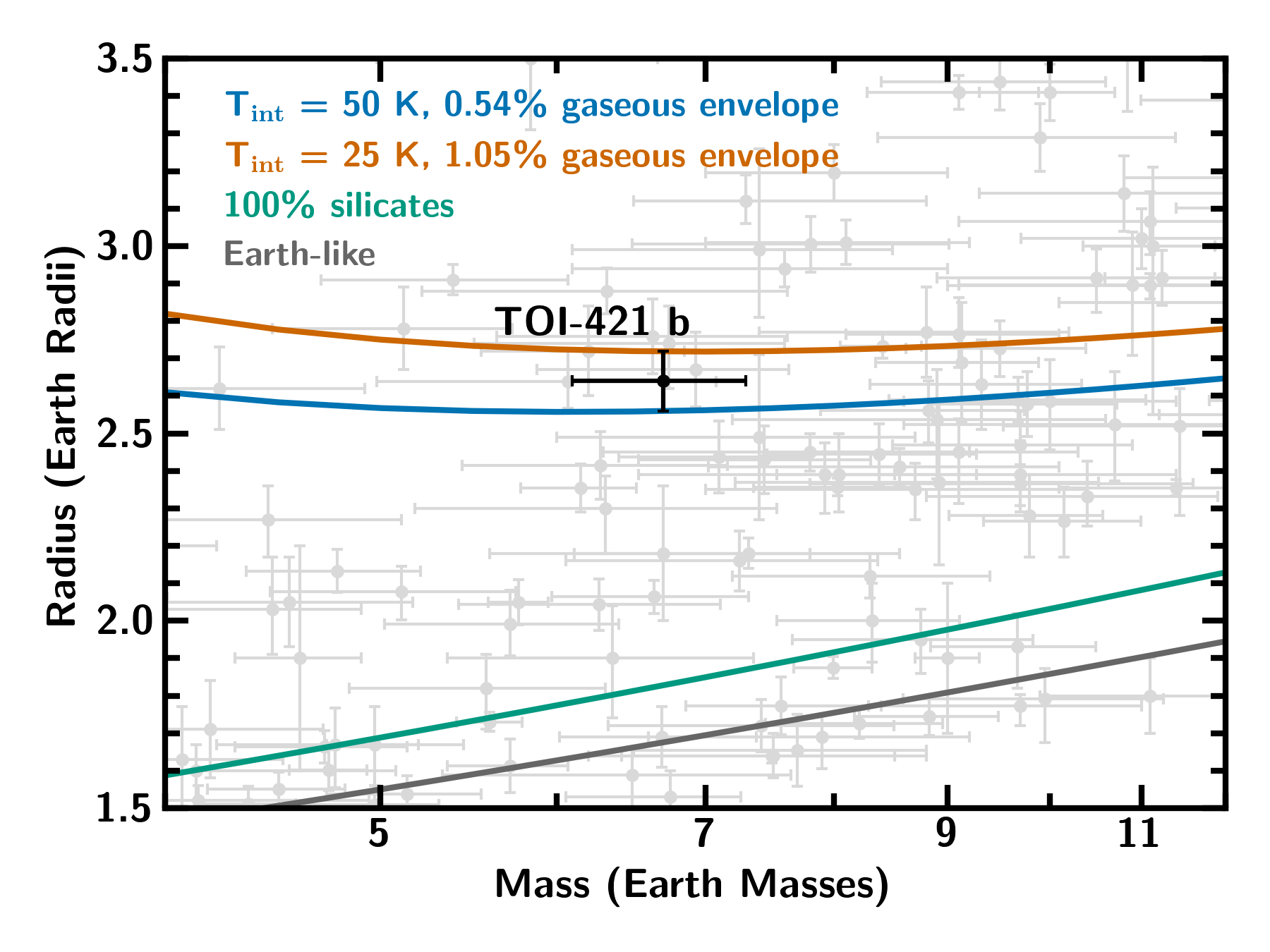}
    \caption{Mass and radius of TOI-421\,b, with mass-radius relations showing the maximum and minimum envelope mass fractions consistent with the planet's measured properties. Percentages shown are the total envelope mass fractions; the envelope itself consists of $\sim$99$\%$ H/He and 1$\%$ H$_2$O by mass, consistent with atmospheric constraints from this study. Earth-like and pure silicate mass-radius relations are shown for comparison. Grey points with error bars show planets with similar masses and radii, taken from the NASA Exoplanet Archive (\url{https://exoplanetarchive.ipac.caltech.edu}).}
    \label{fig:mr_curves}
\end{figure}

We constrain the envelope mass fractions that are consistent with both the atmospheric composition and bulk density of TOI-421\,b. By generating a grid of interior models for both values of $T_{\rm int}$ and a range of envelope mass fractions, we find that the range of envelope mass fractions consistent with the mass and radius of TOI-421\,b to within 1$\sigma$ is 0.54\,--\,1.05\% (see Figure \ref{fig:mr_curves}). The minimum mass fraction is found when $T_{\rm int}$ = 50~K, and the maximum is found when $T_{\rm int}$ = 25~K.

\subsection{Formation and Evolution Implications \label{sec:formation}}
TOI-421\,b's unique stellar environment in comparison to sub-Neptunes previously characterized with JWST may be key to its compositional difference. The results of NASA's Kepler mission, which primarily observed FGK stars, have served as an empirical baseline against which super-Earth and sub-Neptune formation and evolution theories have sought comparison. Specifically, observations show a bimodal distribution of planets with peaks at $\sim$1.3 and $\sim$2.5 $R_{\oplus}$ and a scarcity in between \citep{Fulton2017}. Based on predictions from observationally-derived densities, the two populations are distinguishable as having rocky compositions (super-Earths) or as having thick H/He atmospheres (sub-Neptunes) \citep{weiss14, Rogers2015}, with TOI-421\,b belonging to the latter population.

Formation theories suggest that these differences can follow from similar early formation scenarios (a rocky planetesimal accreting gas from the solar nebula), with the radius gap arising from differences in subsequent atmospheric mass loss. Simply, the now-rocky super-Earths had their primordial envelopes entirely stripped away, while the sub-Neptunes were sufficiently massive to retain atmospheres of up to a few percent of a planet's mass after the early loss. This loss can be driven primarily by photoevaporation due to high-energy flux from young host stars \citep{Owen2017}, or through core-powered mass loss, by which internal luminosity from intensely hot cores drives gas molecules away from the planet \citep{Gupta2019}. In either case, a predominantly H/He envelope is required in order to reproduce the observed properties of the radius valley.  The retrieved low mean molecular weight of TOI-421\,b's atmosphere and its inferred bulk H/He mass fraction of $\sim$1\% from interior modeling are in strong alignment with this picture.

Previous JWST observations do not, however, reveal a set of sub-Neptunes with this simple atmospheric composition. As shown in Figure~\ref{fig:Population} and discussed in Section~\ref{sec:composition}, the other sub-Neptunes observed with JWST to date all have high MMW atmospheres:  GJ 9827\,d is described as a `steam world' \citep{Piaulet2024}; the transmission spectrum of TOI-270\,d spurred the introduction of a new classification of `miscible envelope' sub-Neptunes, in which H/He is well-mixed with a roughly equal mass of metals \citep{Benneke2024}; and K2-18\,b has been alternately described as a Hycean world with an underlying water ocean or as a sub-Neptune hosting a high metallicity atmosphere \citep{Madhusudhan2023, Wogan2024}. Even the planets with featureless spectra have minimum metallicities $>$100$\times$ solar \citep{Gao2023, Wallack2024}. Though these results may represent diversity among a small sample, they may also suggest that the typical M-dwarf sub-Neptune experiences a different formation and evolution history than the canonical Kepler radius valley planets. In contrast, TOI-421\,b, the first sub-Neptune around a FGK star to be characterized with JWST, does host a hydrogen-dominated envelope, in support of theories that mass loss governs the formation of the radius valley.

The simple picture of sub-Neptunes being composed of primordial H/He atmospheres overlying rock/iron cores may additionally break down when considering other physical processes that would be expected to further alter their atmospheric composition.  For example, sub-Neptunes are expected to have long-lived magma oceans, which through dissolution and outgassing at the magma-atmosphere interface can reprocess the atmospheric composition until it no longer strongly represents its conditions at formation \citep[e.g.,][]{chachan18,kite20}.  Depending on whether the planet's iron core interacts chemically with the mantle and atmosphere, large quantities of gaseous H$_2$O can be generated through interactions between molten silicates and hydrogen gas, producing highly water-enriched atmospheres \citep{schlichting22,Seo2024}.  Photoevaporative mass loss may additionally shape atmospheric composition via fractionation that can occur in the outflowing gas.  To first order, mass loss should enhance metallicity and reduce C/O: hydrogen is lost readily due to its low mass, and carbon's low ionization potential means that it is more easily entrained in the photevaporative outflow than oxygen \citep{owen22,Seo2024}.  We see no direct evidence for such processes impacting the composition of TOI-421\,b's atmosphere: our measurements are entirely consistent with a truly primordial envelope. However, we are limited by our inability to precisely constrain elemental abundance ratios such as C/O, which would be more revealing as to the role of magma ocean interactions and fractionated mass loss in sculpting the atmosphere of this low-metallicity sub-Neptune.

\subsection{Observing Mode Considerations}

In this work, we present one of the first published results of an exoplanet observation with NIRSpec/G395M.  Benefits of the G395M mode (as opposed to the more widely used G395H) are slightly higher throughput and that the entire observation sits on a single detector, negating any concerns about a gap in spectral coverage or the potential for offsets in transit depths between detectors. The downsides of the G395M mode are a reduction in spectral resolution and faster saturation, which requires a sacrifice in up-the-ramp sampling that may lead to additional noise due to fewer groups for ramp fitting.

We find that our ability to detect molecules absorbing in the 4\,--\,5 $\mu$m region is reduction-dependent. The wavelength bins that disagree to near or greater than 1$\sigma$ are sufficient to influence the detection significance of SO$_2$ and CO for any given retrieval model setup, as shown in Table \ref{tab:results}. We also note that the values of retrieved atmospheric parameters can be sensitive to differences in transit depths of $\leq1\sigma$ between reductions \citep[e.g.,][]{Lewis2020}. We performed a step-by-step comparison of our independent NIRSpec reductions including, but not limited to, choices in which stage of data to start with (e.g., \texttt{uncal} vs.\ \texttt{rateints}), spectral extraction (e.g., optimal vs.\ standard), outlier removal on the light curves themselves, and light curve fitting methods (e.g., MCMC vs.\ least squares), and found that no single choice in the handling of systematics dominates the difference in results. The relatively small feature sizes for this data set compared to the noise level of the detector is the most likely reason for the different results depending on reduction choices. Additional data would help to improve agreement and overcome the noise sources.

Importantly, the white light curves obtained from these observations do not reach the level of precision seen in NIRSpec/G395H data sets. Our \texttt{Eureka!}\ NIRSpec/G395M reduction achieves at best $\sim$1.7$\times$ photon noise, while it is common to reach closer to $\sim$1.2$\times$ photon noise for NIRSpec/G395H \cite[e.g.,][]{Lustig-Yaeger2023}. The reason for this difference is not immediately clear, though it may be related to the fact that the G395M mode only illuminates the NRS1 NIRSpec detector (albeit on different pixels from the G395H mode), whereas NIRSpec/G395H disperses light over both detectors. The NRS2 detector has been observed to have less systematic noise than NRS1 \citep[e.g.,][]{Wallack2024,luque24}.  Further work should explore the relative capabilities of these modes.

The NIRSpec/G395 observing modes have been favored by many exoplanet atmosphere programs with JWST due to their coverage of key carbon, oxygen, and sulfur-bearing species.  Despite our challenges with the noise properties of the G395M mode, our NIRSpec transit observation contains valuable information. The short-wavelength water slope, the poor fit to models with high metallicity and with high C/O from methane, and even the tentative detections of SO$_2$ and CO all point to the fact that TOI-421\,b is unique among sub-Neptunes observed to date and that it is a prime candidate for further in-depth characterization. 

However, we also find that the NIRISS/SOSS measurements alone are effective for measuring the atmospheric water abundance and constraining its MMW (see Table \ref{tab:results}). This agrees with the work of  \cite{Piaulet2024}, who found that NIRISS combined with HST/WFC3 enabled the measurement of water in the warm sub-Neptune GJ 9827\,d, even in the high metallicity regime.   We further note that, had we only observed with NIRSpec/G395M, we would have incorrectly concluded that the transmission spectrum of TOI-421\,b is featureless.  We point this out as a cautionary tale and as evidence for the usefulness of the NIRISS/SOSS observing mode in the characterization of sub-Neptune atmospheres.

\section{Conclusion}  \label{sec:conclusion}
We analyzed the JWST transmission spectrum for TOI-421\,b, the first for a hot sub-Neptune around an FGK star. In doing so, we find:
\begin{itemize}
    \item Evidence for a cloud- and haze- free upper atmosphere, supporting evidence that photochemical hazes are unlikely to be present in hot sub-Neptune atmospheres.
    \item A robust detection of H$_2$O, along with tentative detections of SO$_2$ and CO, which hint at the presence of photochemistry and demonstrate the dominance of CO/CO$_2$ over CH$_4$ in warmer atmospheres.
    \item A low-MMW atmosphere, indicative of a hydrogen-dominated envelope at near-solar metallicity, in contrast with recent measurements of cooler sub-Neptunes around M-dwarf stars.
\end{itemize}
These findings, along with our inferred bulk $\sim$1\% H/He mass fraction, imply that TOI-421\,b hosts a primordial atmosphere, in line with predictions that the radius valley is shaped by mass loss processes.  Furthermore, we do not find obvious evidence for additional processes such as magma ocean interactions or fractionated outflows playing a prominent role in establishing the atmospheric composition of this planet, despite predictions that would imply otherwise.

The tantalizing differences between the properties of TOI-421\,b's atmosphere and those of other JWST-observed sub-Neptunes orbiting M-dwarf stars indicate the need for further study of objects in this class.  In particular, future work should focus on whether TOI-421\,b's properties are emblematic of hot sub-Neptunes orbiting FK stars or whether the planet's properties simply reveal further compositional diversity among the sub-Neptune population.  Two specific questions deserve further attention:  `Do all sub-Neptunes hotter than $\sim$850~K have aerosol-free atmospheres?', and `Do sub-Neptunes orbiting FGK-type hosts typically have low-metallicity envelopes?'.  Improved data for TOI-421\,b and similar sub-Neptunes in the \mbox{3\,--\,5 $\mu$m} spectral range are also vital for constraining their carbon and sulfur reservoirs and measuring elemental abundance ratios. These are needed to better inform formation and evolution conditions along with physical processes such as magma ocean interactions and mass loss that sculpt the observed properties of these planets' atmospheres.


\section*{Acknowledgements}
This work is based on observations made with the NASA/ESA/CSA James Webb Space Telescope. The data were obtained from the Mikulski Archive for Space Telescopes at the Space Telescope Science Institute, which is operated by the Association of Universities for Research in Astronomy, Inc., under NASA contract NAS 5-03127 for JWST. These observations are associated with program \#1935. Support for this program was provided by NASA through a grant from the Space Telescope Science Institute. This work benefited from the 2024 Exoplanet Summer Program in the
Other Worlds Laboratory (OWL) at the University of California, Santa Cruz,
a program funded by the Heising-Simons Foundation.
We thank Michael Radica for his help implementing the \texttt{exoTEDRF} pipeline and for providing insight into the NIRISS/SOSS observing mode.


\section*{Data Availability}

The spectroscopic and white light curves derived in this paper as well as posterior distributions for all retrievals listed in Table~\ref{tab:results} will be made available on Zenodo upon acceptance of this paper for publication.


\bibliography{REFS}{}
\bibliographystyle{aasjournal}
\end{document}